# The TESLA Requirements Database

L. Hagge, J. Kreutzkamp, K. Lappe
*Deutsches Elektronen-Synchrotron (DESY), Hamburg, Germany*

In preparation for the planned linear collider TESLA, DESY is designing the required buildings and facilities. The accelerator and infrastructure components have to be allocated to buildings, and their required areas for installation, operation and maintenance have to be determined. Interdisciplinary working groups specify the project from different viewpoints and need to develop a common vision as a precondition for an optimal solution. A commercial requirements database is used as a collaborative tool, enabling concurrent requirements specification by independent working groups. The requirements database ensures long term storage and availability of the emerging knowledge, and it offers a central platform for communication which is available for all project members. It is successfully operating since summer 2002 and has since then become an important tool for the design team.

## 1. INTRODUCTION

Planning a new accelerator involves among other designing the machine, performing civil engineering, planning the layout of technical infrastructure, and handling public affairs. The planning team includes scientists, engineers and technicians from accelerator and research groups, the survey and safety departments, cryogenics, HF, electricity providers and many more. Coordinating the planning team and the planning activities is a challenging task for project management.

### 1.1. Coordinating Planning Activities

The goal of the activities is to design the buildings and facilities which are required for the planned linear accelerator TESLA [1] [2], and to describe the major procedures for construction, installation, operation and maintenance. Several planning documents need to be created for the different purposes: Architectural models specify the layout of the various tunnels and buildings, building plans contain mock-ups and cross sections of the different buildings, construction plans describe the construction and installation procedures, etc.

The planning team has been organized into several distributed expert groups for specific topics. The groups perform their tasks independently, but they have to develop a common vision in order to be able to create a joint design of the accelerator and its facilities.

In the beginning, each group starts from its professional point of view. As work proceeds, the specifications begin to overlap. For example, scientists require the experimental hall to be as large as possible for the detector and all its supplies, while management and public relations require the hall to be as small as possible to reduce cost and environmental impact. Figure 1 summarizes some of the planning documents and illustrates that each document is created by a group of experts, and experts might contribute to more than one group. It also shows that the documents depend on each other: If e.g. the architectural model is changed, the building plans have to be regenerated, the construction procedure might have to be adapted, and it has to be ensured that the modifications are compatible with the safety concept.

In addition to the dependencies among the documents, the different facilities also depend on each other at a technical level. Changing e.g. the parameters of a detector or a cryogenic plant implies also changes in the requirements on supporting buildings and installations. The authoring expert groups thus need to exchange information on requirements and boundary conditions with other groups at different locations in order to proceed with their work.

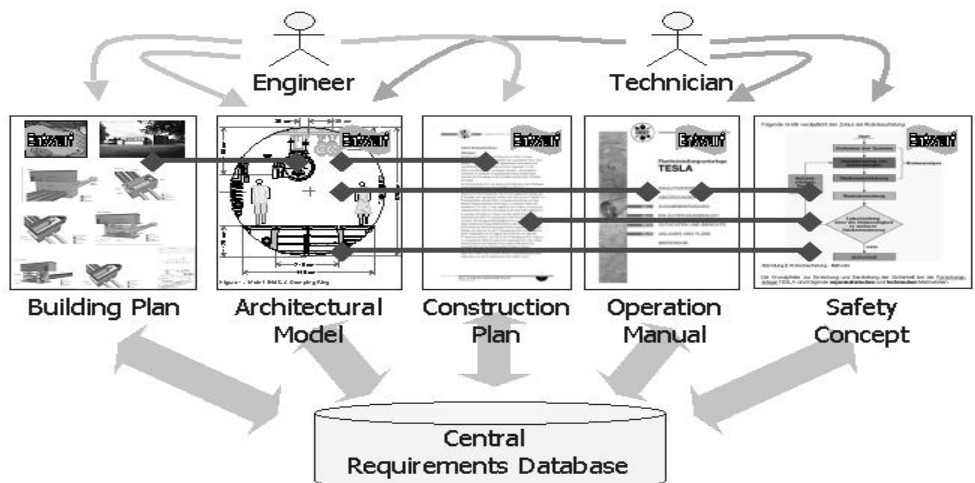

Figure 1. Planning documents and their dependencies.

## 1.2. Requirements Engineering

Establishing close collaboration among the different expert groups implies that each planning decision has to be documented and communicated among the entire planning team, and impact analysis on other work in progress has to be performed. What is needed is an efficient toolset for creating documentation and for classifying, tracking, analyzing and reporting every single documentation item, together with a set of rules for specification and tool usage.

Requirements Engineering (RE, e.g. [3], [4]) has been introduced to support the distributed expert groups in requirements elicitation, negotiation, documentation and verification. Particular goals included

- enabling concurrent specification of components by several independent expert groups,
- providing methods and criteria for requirements negotiation, validation and approval,
- providing a central communication and documentation platform which is available to the entire planning team, and
- ensuring long term archiving and availability of emerging knowledge.

## 2. REQUIREMENTS MANAGEMENT ARCHITECTURE

The described RE solution is based on a commercial requirements management system (RMS [4]) which is used as a collaborative tool. The RMS enables concurrent requirements specification by the expert groups and provides a single point of access for information retrieval. The following sections briefly describe the basic procedures for requirements specification, requirements retrieval and requirements publication.

### 2.1. Requirements Specification

Every expert group had to nominate one person who became responsible for requirements specification and documentation (GRM, group requirements manager). The GRM had to lead the discussion within the teams, write the requirements specification documents, and find and resolve conflicting requirements from different expert groups.

The GRMs create requirements specification documents using their accustomed office word processors. The office tools are connected with the RMS, which stores individual paragraphs from the documents as requirements in its central repository. The specification documents from different expert groups are kept in separate documents. This way, the documents can be modified independently, thus enabling distributed specification. Figure 2 illustrates how different expert groups maintain separate specification documents, which are logically connected within the RMS.

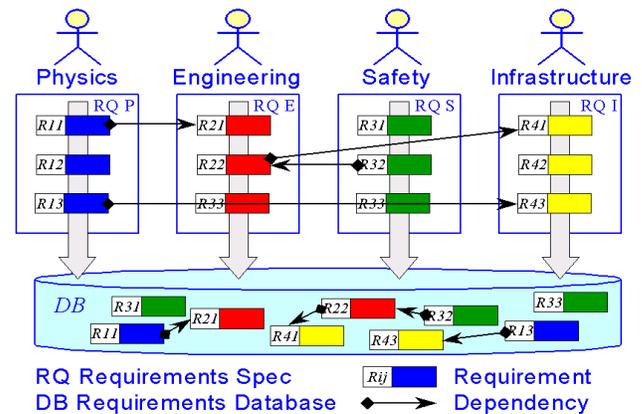

Figure 2. Distributed requirements specification.

The RMS maintains metadata which allow to store further details for each requirement. Some project-specific metadata attributes have been configured in the RMS, including the requirements type, the authoring group and the affected buildings and sites. In addition, the RMS stores default metadata like e.g. version counters, dates of change and a change log. Figure 3 illustrates a requirement data record in the RMS.

```
During installation, consoles at beam
height are needed in the experimental         R 234
hall for measuring.

type      =     techn. infrastructure,
                floor space.
group     =     survey
building  =     experimental hall
location  =     site-01
phase     =     installation
....
```

Figure 3. Example for requirement data record.

It is the responsibility of the GRMs to provide correct metadata information for each requirement. To ease using the RMS, the attributes are configured to offer value lists. For example, the requirement's type is chosen from "usage", "technical infrastructure", "floor space" or several non-functional constraints like "safety", "cost" etc. High-quality metadata is essential for efficient requirements retrieval.

### 2.2. Requirements Retrieval

Requirements are mainly retrieved using database queries. This way, requirements originating from different groups but addressing the same topic can be jointly found and retrieved.

Figure 4 shows an example of a civil engineer searching the database for all floor space requirements on the experimental hall. The result yields the floor space requests from all the expert groups and serves as design specification for the hall.

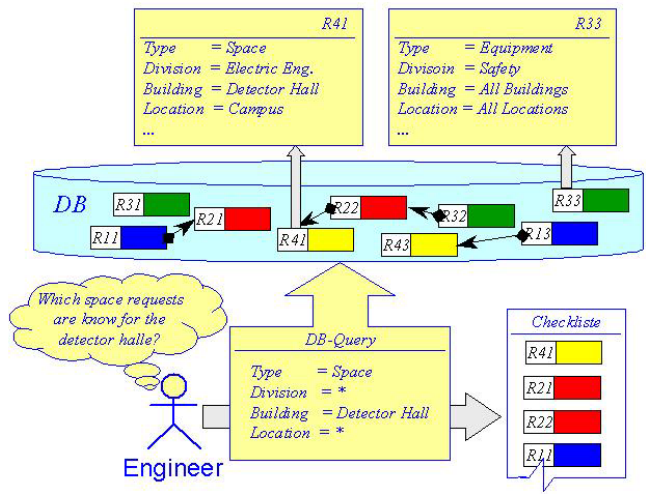

Figure 4. Requirements retrieval.

Most of the design specifications for the buildings are generated in a similar way, and it has to be emphasized that a single authoring group would probably have had difficulties to manually assemble an equally complete specification.

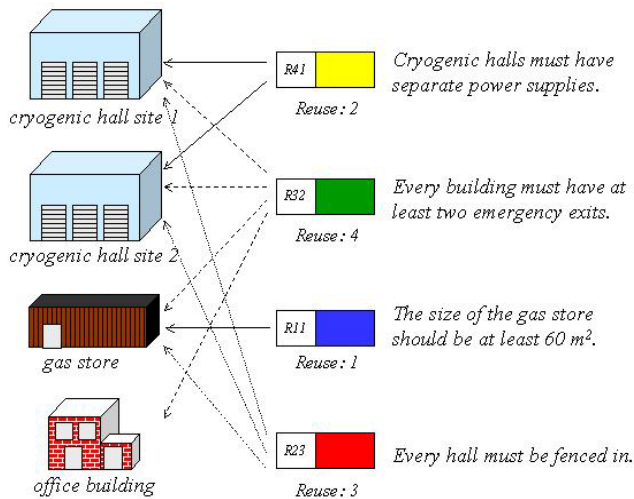

Figure 5. Requirements reuse.

A very powerful feature of the RMS is the possibility of reusing requirements. Multiple-value attributes allow to assign requirements to several locations or buildings. For example, the mandatory availability of emergency exits in every building can be specified easily by entering the requirement into the database only once, and then selecting all the buildings in the corresponding attribute. The requirement will from then on appear in the design specification for every building. Figure 5 illustrates some examples for reuse of requirements.

### 2.3. Requirements Presentation

The RMS can be accessed through a powerful Web-interface which is available to the entire planning team.

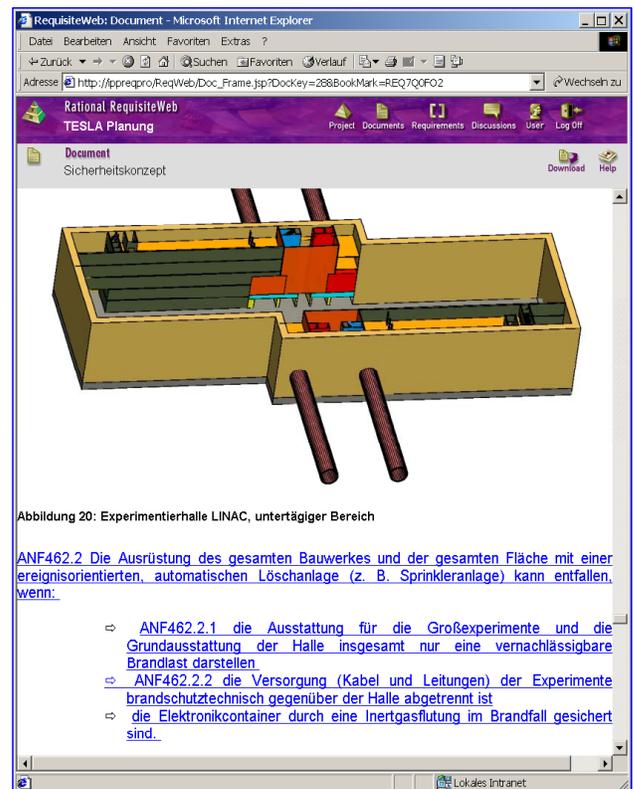

Figure 6. Example for a requirements specification document on the Web.

Apart from creating specification documents, the Web interface offers the full functionality of the RMS. This includes in particular

- viewing the requirements specification documents of all expert groups in HTML (cf. Figure 6),
- searching the database, creating ad-hoc queries and storing views,
- filtering the database using predefined views (cf. Figure 7), and
- navigating from documents to requirements data records and back.

Figure 7. Example for a filtered database view on the Web.

*Computing in High Energy and Nuclear Physics CHEP2003, UC San Diego, March 24-28, 2003*

## 3. RESULTS

The TESLA planning team is using the RMS since summer 2002. The following two examples highlight the benefits which were experienced by the team.

### 3.1. Discovering Dependencies

In a review meeting in an early project stage, the RMS was queried explicitly for requirements which were affecting the experimental hall, but had been specified outside the responsible expert group. It turned out that the survey department had specified consoles at beam height to accommodate their equipment, but that the information had not yet been passed to the design engineers and were thus not included in the design. The necessity of the consoles and the space requirements were then negotiated.

Querying the database helps to identify conflicting requirements or omissions early in the project, the expert groups simply have to search for entries from different authors which affect their topics of interest. This way, the RMS fosters communication among the expert groups and helps to improve the quality of the specification and the resulting design.

### 3.2. Generating Approval Checklists

In later project stages the resulting designs need to be approved and frozen. The responsible project engineers usually do not know all the requirements which contribute e.g. to a certain building. The RMS enables generating checklists from the database, which contain the entire specification and can then be used to verify if a building design fulfills all its requirements (cf. Figure 8). The results of the approval can be read back into the RMS, thus enabling the entire planning team to easily identify at any which requirements are already fulfilled, and which topics still need discussion.

| Checklist Experimental Hall | | |
|---|---|---|
| A storage room for electrical equipment of about 80 m² is needed. | R 123 | ✓ |
| It must be accessible by car. | R 123.1 | ✓ |
| During installation, in the experimental hall consoles at beam height are needed for measuring. | R 234 | ⊘ |
| Consoles must be accessible by gangways. | R 234.1 | ⊘ |
| ... | | |

Figure 8. Example for an approval checklist.

As the RMS traces the history of each requirement, it is also possible to identify which requirements have been modified after they have already been used for the approval of e.g. a certain building design.

## 4. EXPERIENCE

Requirements continue to evolve as the planning activities proceed, thus they have to be managed and made available in updated versions. Requirements negotiation and the extraction of valid specifications is an essential task in each project, which can be successfully supported and managed by an RMS.

The RMS is only a tool, but projects need to adopt an RE method. In the described project, the RMS was the means to introduce and establish requirements engineering in the planning team. The success and acceptance of the RMS were mainly due to the following key factors:

- The responsibility for was clearly defined by nominating a requirements manager in every group.
- Technical and methodical support staff backed up the introduction of requirements engineering. They assisted in creating specifications and provided support and training for the GRMs (cf. Figure 9).
- The RMS tool was easy to use and provided information which was nowhere else available .
- The requirements were necessary for approval procedures of building designs, which ensured back-up and support by project management.

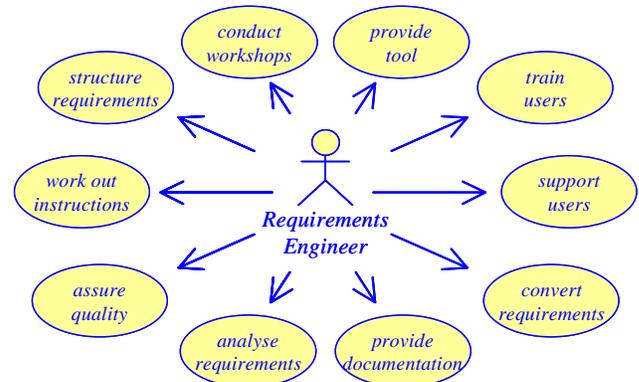

Figure 9. Responsibilities of a requirements engineer.